\documentclass[twocolumn,superscriptaddress]{revtex4}
\usepackage{amssymb}
\usepackage{amsmath}
\usepackage{graphicx}
\usepackage{color}

\begin{document}

\title{Well-tempered metadynamics: a smoothly-converging and tunable
free-energy method}
\author{Alessandro Barducci}
\affiliation{Museo Storico della Fisica e Centro Studi e Ricerche Enrico Fermi, Compendio Viminale, 00184 Roma, Italy}
\affiliation{Istituto Nazionale di Fisica Nucleare, Sezione di Roma Tor Vergata, Via della Ricerca Scientifica 1, 00133 Roma, Italy}
\affiliation{Computational Science, Department of Chemistry and
Applied Biosciences, ETH Z\"urich, USI Campus, Via Giuseppe Buffi 13,
CH-6900 Lugano, Switzerland}
\author{Giovanni Bussi}
\email{gbussi@ethz.ch}
\author{Michele Parrinello}
\affiliation{Computational Science, Department of Chemistry and
Applied Biosciences, ETH Z\"urich, USI Campus, Via Giuseppe Buffi 13,
CH-6900 Lugano, Switzerland}

\begin{abstract}
We present a method for determining the free energy dependence on a
selected number of collective variables using an adaptive bias.
The formalism provides a unified
description which has metadynamics and canonical sampling as limiting
cases. Convergence and errors can be rigorously and easily
controlled.
The parameters of the simulation can be tuned so as
to focus the computational effort only on the physically relevant
regions of the order parameter space. The algorithm is tested on the
reconstruction of alanine dipeptide free energy landscape.
\end{abstract}

\maketitle

Computing free energy differences is of crucial importance in
molecular dynamics (MD) and Monte Carlo (MC) simulations.  Whenever it
is possible to define a few collective variables (CVs) that provide
a
coarse-grained description of the slow modes \cite{gear+02cce,humm-kevr03jcp},
it is also of great relevance to
compute the associated free energy surface (FES).  In order to draw
such a surface a 
straightforward 
approach is often not
possible due to 
high barriers or other sampling
bottlenecks.  A standard strategy for overcoming this problem is to
introduce an external biasing potential that forces the system to
explore regions of high free energy \cite{torri-valle77jcp}.  A major
progress has been the recent introduction of adaptive non-equilibrium
methods
\cite{wang-land01prl,darv-poho01jcp,laio-parr02pnas,mars+06jpcb}.
In all these methods the simulation history is used to enhance
the sampling speed.  In a MC run, this can be done by
varying the MC acceptance probability every time a new configuration
is visited \cite{wang-land01prl}, while in MD a time-dependent bias
can be added either to the force \cite{darv-poho01jcp} or to the
potential \cite{laio-parr02pnas,mars+06jpcb}.

We shall focus here on metadynamics \cite{laio-parr02pnas}, which has
proven its effectiveness in a variety of contexts
\cite{mich+04prl,gerv+05jacs,ensi-klei05pnas,ogan+05nature,ishi+06prl,nair+06jacs,boer+06jacs,%
buss+06jacs,kuma+07jcp,spiw+07jpcb,lee+06jpcb}.  In metadynamics the
system evolution is biased by a history-dependent potential that is
constructed as the sum of Gaussian functions \cite{hube+94jcamd}
deposited along the trajectory in the CVs space.  After a transient,
the bias potential compensates the underlying FES and provides
an estimate of its dependence on the CVs. 
A formal justification of this procedure has been given in Ref.~\cite{buss+06prl}. 
In spite of its success there is a need to improve metadynamics in
several respects. First of all, it is often difficult to decide when
to terminate a metadynamics run.  In fact, in a single run, the free
energy does not converge to a definite value but
fluctuates around the correct result, leading to an average error which is
proportional to the square root of the bias potential deposition rate
\cite{laio+05jpcb,buss+06prl}. Reducing this rate implies increasing
the time required to fill the FES. Furthermore,
in practical application, continuing               a  run
carries the risk that the system is irreversibly pushed in regions of
configurational space which are not physically relevant.  These issues
have already been recognized and different ad-hoc solutions have been
proposed to alleviate these problems
\cite{mich+04prl,wu+04jcp,gerv+05jacs,ensi-klei05pnas,babi+06jcp,min+07jcp}.

In this Letter, inspired by the self-healing umbrella sampling method
\cite{mars+06jpcb}, we substantially improve metadynamics such
that we obtain an estimate of the FES that
converges to the exact result in the long time limit.
Contrary to ordinary metadynamics, our approach offers the possibility
of controlling the regions of FES that are physically meaningful to
explore.  Besides being highly effective and controllable this new
method provides a unified framework whose limiting cases are standard
metadynamics and non-biased standard sampling.
We dub this new scheme well-tempered metadynamics.

Let us consider a system described by a set of microscopic
coordinates $q$ and a potential energy $U(q)$, evolving under the
action of a dynamics (e.g. MD or MC)
whose equilibrium distribution is canonical at the temperature
$T$.
We want to determine the free energy dependence on a set
of collective variable $s(q)$.
The FES can be written within an immaterial constant as 
\begin{equation} \label{log}
F(s)=-T \lim_{t\rightarrow\infty} \ln N(s,t),
\end{equation}
where $N(s,t)=\int_{0}^{t} \delta_{s,s(t')} dt'$ is 
the histogram of the variable $s$ obtained from an unbiased simulation.
By construction, $N(s,0)=0$ and its time
derivative $\dot{N}(s,t)=\delta_{s,s(t)}$.
To accelerate sampling we bias the dynamics by
adding the history-dependent potential
\begin{equation} \label{master-equation}
V(s,t)=\Delta T \ln \left(
  1+\frac{\omega N(s,t)}{\Delta T}
\right),
\end{equation}
where $\omega$ has the dimension of an energy rate, $\Delta T$ is a
temperature and $N(s,t)$ comes from the biased simulation.  Since $V$
is a monotonic function of $N$, such a bias potential disfavors the
more frequently visited configurations.
A crucial quantity 
is the rate at which the potential is modified. In particular, slower
variation rates lead to a dynamics of the microscopic variables $q$
which is closer to thermodynamic equilibrium. From
Eq.~(\ref{master-equation}) it follows that the rate with which
$V(s,t)$ changes is:
\begin{equation} 
\label{dotv}
\dot{V}(s,t)=\frac{\omega \Delta T \delta_{s,s(t)}}{\Delta T+\omega N(s,t)} = 
\omega e^{-\frac{V(s,t)}{\Delta T}} \delta_{s,s(t)}.
\end{equation}
The connection with metadynamics is evident if we examine
Eq.~(\ref{dotv}) and replace $\delta_{s,s(t)}$ with a finite width
Gaussian.
Therefore, our scheme can easily be implemented in any metadynamics code
by rescaling the height of the Gaussians
according to Eq.~(\ref{dotv}).
Using the notation in Ref.~\cite{laio+05jpcb},
the height of each Gaussian is determined by
$w=\omega e^{-\frac{V(s,t)}{\Delta T}} \tau_{\text{G}}$,
where $\tau_{\text{G}}$ is the time interval at which Gaussians are deposited.
Thus $\omega$ represents the initial bias deposition rate.

Two important properties need to be
underlined. The first is that since the histogram $N(s,t)$ grows
linearly with simulation time, the rate $\dot{V}(s,t)$ tends to zero as
$\propto 1/t$. This is the simplest, if possibly not the optimal
\cite{poul+06pre}, way to have a rate decrease fast enough for the
bias eventually to converge, yet slow enough for the final result not
to depend on the initial condition $V(s,0)$.  Similar arguments have
been used in the field of stochastic optimization
\cite{harj+97prl,spal03book}.  The second property is that $\dot{V}$
is not uniform in the $s$ space since at a given point the rate is
inversely proportional to the time already spent there. This latter
feature distinguishes our approach from others in which $1/t$ strategy
has also been suggested either explicitly \cite{bela-pere07pre} or
implicitly \cite{mars+06jpcb}.

For large times, $V(s,t)$ varies so slowly that one can assume that
the $q$'s reach equilibrium, the probability distribution becomes
$P(s,t)ds\propto \exp\left(\frac{-F(s)-V(s,t)}{T}\right)ds$ and one
has:
\begin{equation}
\label{limit-dotv}
\dot{V}(s,t)=\omega e^{-\frac{V(s,t)}{\Delta T}}P(s,t)= \omega
e^{-\frac{V(s,t)}{\Delta T}}\frac{e^{-\frac{F(s)+V(s,t)}{T}}}{\int ds ~
e^{-\frac{F(s)+V(s,t)}{T}}}.
\end{equation}
This implies that $V(s,t\rightarrow\infty) =-\frac{\Delta T}{\Delta
T+T}F(s)$, modulo a constant. Thus at variance with metadynamics and
other methods the bias does not fully compensate $F(s)$, rather one
has that $F(s)+V(s)=\frac{T}{T+\Delta T}F(s)$ leading to the following
distribution of $s$:
\begin{equation}
\label{limit-p-of-s}
P(s,t\rightarrow\infty)ds\propto e^{ -\frac{F(s)}{T+\Delta T}}ds.
\end{equation}
In practice, using Eq.~(\ref{master-equation}) the FES can be estimated as
\begin{equation}\label{estimate}
\tilde{F}(s,t)
=-\frac{T+\Delta T}{\Delta T} V(s,t)
=-(T+\Delta T) \ln\left(
1+\frac{\omega N(s,t)}{\Delta T}
\right).
\end{equation} 

Let us examine the two limiting cases,
$\Delta T=0$ and $\Delta T\rightarrow\infty$.  For $\Delta T=0$ the
bias is equal to zero and Eq.~(\ref{estimate}) reduces to
Eq.~(\ref{log}).  More interesting is the $\Delta T\rightarrow\infty$
limit.  In this case, the deposition rate is constant, and from
Eq.~(\ref{estimate}) one finds that $\tilde{F}(s,t)=-V(s,t)$ and the
standard metadynamics algorithm is recovered.  Note however that the
limit $\Delta T\rightarrow\infty$ is singular: if we first let $\Delta
T\rightarrow\infty$, the convergence of $V(s,t)$ for
$t\rightarrow\infty$ cannot be demonstrated by means of
Eq.~(\ref{limit-dotv}). This is a reflection of the already noted
drawback of metadynamics that in a single simulation, the bias
does not converge but oscillates around the correct
$F(s)$ value.
In intermediate cases the calculated FES is the one corresponding to
the target temperature $T$, with the transverse degrees of freedom
correctly sampled.  However, the $s$ probability distribution is
altered and corresponds to an enhanced temperature $T+\Delta T$. It
must be stressed that this result has been obtained without having to
assume adiabatic separation between $s$ and the other variables 
as in Refs.~\cite{vand-roth02jcp,ross+02jcp,mara-vand06cpl}.

Much is to be gained
computationally by well-tempered metadynamics.  By tuning $\Delta T$
one can increase barrier crossing and facilitate the exploration in
the CVs space. Furthermore using a finite value of $\Delta T$ one
automatically limits the exploration of the FES region to an energy
range of the order $T+\Delta T$. Hence the
exploration of the FES can be limited to the physically interesting
regions of $s$.  Longer simulation time results in
improved statistical accuracy in the relevant regions.  The risk of
overfilling is avoided and optimal use is made of the computer time.
Deciding when to stop the run is now simple and
post-processing \cite{wu+04jcp,mich+04prl} is not necessary.

As an illustration we study the FES of alanine
dipeptide in vacuum as a function of the backbone dihedral angles
$(\Phi,\Psi)$.  This surface has been well studied and is known to
exhibit two minima $C_{7eq}$ and $C_{7ax}$ separated by a barrier of
$\approx$9 kcal mol$^{-1}$ \cite{mara+06jcp,bran+07jcp}.  Since such a barrier cannot be crossed with
standard dynamics at room temperature this system has provided a
testing ground for many sampling schemes.  The CHARMM27 \cite{mack+98jpcb}
force field has been used in ORAC MD code \cite{ORAC}
and canonical sampling at a temperature of 300 K was achieved by means
of the stochastic thermostat in Ref.~\cite{buss+07jcp}.  The
Gaussian width was set to 20 degrees, and the deposition interval was 120
fs with a starting Gaussian height of 0.287 kcal mol$^{-1}$,
which corresponds to
a deposition rate $\omega$=2.4 cal mol$^{-1}$fs$^{-1}$.

\begin{figure}
\includegraphics[clip,width=0.45\textwidth]{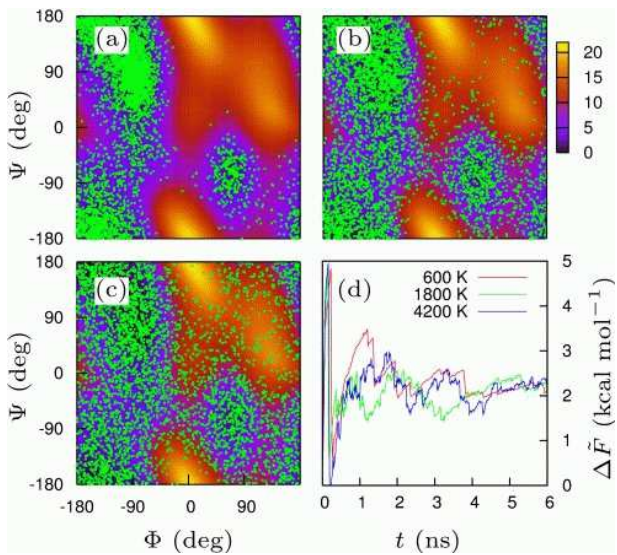}
\caption{
\label{fig1}
Panels (a, b, c): Green dots represent 6 ns long
trajectories in the $(\Phi, \Psi)$ space for different choices of
$\Delta T$ [600 K(a), 1800 K (b), and 4200 K (c)]. The underlying
color map (kcal mol$^{-1}$) shows the reference free energy landscape.
Panel (d): Estimate of the free energy difference
between the two metastable minima $C_{7ax}$ (70,-70) and $C_{7eq}$
(-83,74) as a function of the simulation time, as obtained from the same trajectories.
}
\end{figure}

We calculated
a reference $F(\Phi,\Psi)$ using standard umbrella sampling which is
in good agreement with previous studies.
On this surface we superimpose three different
trajectories (see Fig.~\ref{fig1}) started from the same initial conditions [$C_{7eq}$ (-83,74)], but with three different choices of
$\Delta T$ (600 K, 1800 K and 4200 K).
In all three cases the secondary metastable state
$ C_{7ax}=(70,-70)$ was frequently visited.
It is worth noting that, as discussed earlier, by increasing
$\Delta T$ larger and larger regions were explored.
In order to demonstrate how the method converges,
for the three mentioned cases, in Fig.~\ref{fig1} we also show
the time evolution of $\Delta \tilde{F}(t) =
\tilde{F}(C_{7ax},t)-\tilde{F}(C_{7eq},t)$, {\it i.~e.~}the estimated free
energy difference between the two minima.
$\Delta \tilde{F}(t)$ converges to the reference value
($\Delta F\approx$2.2 kcal mol$^{-1}$)
in all three trajectories. At variance with standard metadynamics,
the time derivative of the bias potential tends to zero and
the fluctuations around the correct value are progressively damped.
All three simulations provide an accurate estimate of the free energy
difference within a few nanoseconds, even in the lowest  $\Delta T$ case
where the lower number of barrier crossing events leads to a
jumpier $\Delta \tilde{F}$ evolution.
 
\begin{figure}
\includegraphics[clip,width=0.45\textwidth]{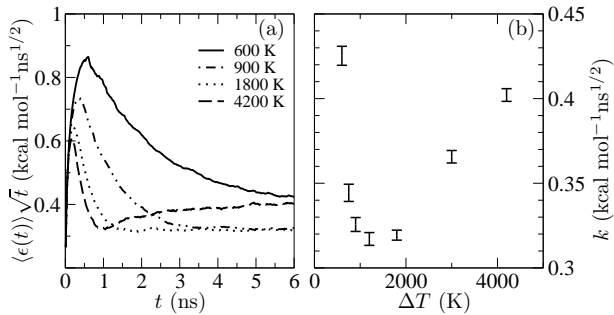}
\caption{
\label{figerror}
Panel (a): Time evolution of $\langle\epsilon(t)\rangle\sqrt{t}$ for
different choices of $\Delta T$.  $\langle\epsilon(t)\rangle$  is the error as defined in Eq.~(\ref{eq:error}), averaged
over an ensemble of 100 independent atomistic simulations starting from
$C_{7eq}$. Panel (b): Dependence of $k$ (see text for definition)
on $\Delta T$, as estimated from 6 ns long trajectories.
}
\end{figure}
\begin{figure}
\includegraphics[clip,width=0.45\textwidth]{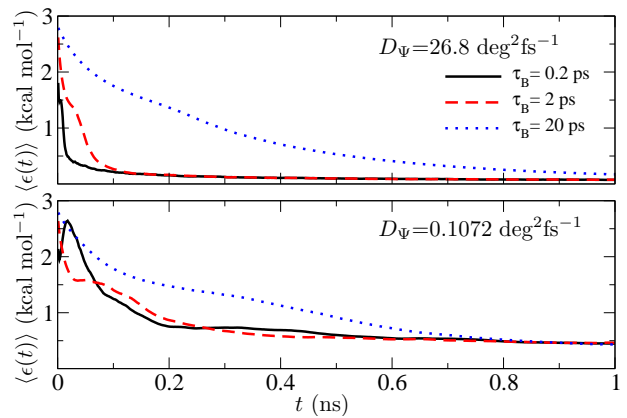}
\caption{
\label{figmodel}
(color online) Time evolution of the average error $\langle\epsilon(t)\rangle$
for different values of $\tau_{\text {B}}$ and $D_{\Psi}$, where
$D_{\Phi}=12.3$ deg$^2$fs$^{-1}$ and $\Delta T$=1200 K.
The error is averaged over an ensemble of 1000 independent Langevin
simulations starting from $C_{7eq}$. }
\end{figure}

As a measure of the error of $\tilde{F}(\Psi,\Phi)$ in the relevant regions,
we define
\begin{equation}
\label{eq:error}
\epsilon(t)= \left(\frac{1}{A} \int_{\Gamma}(F(\Psi,\Phi)-\tilde{F}(\Psi,\Phi,t)-C(t))^{2}d\Phi d\Psi \right)^{1/2}
\end{equation}
where $\Gamma$ is the region in dihedral space such that
$(F(\Psi,\Phi)-F(C_{7eq}))<10$ kcal, and $A$ is its area. 
$\Gamma$ is defined to include all the minima and all the transition states. 
The value of $C(t)$ is
chosen so as to align the averages of $F$ and $\tilde{F}$ over $\Gamma$.
It is seen that after an initial transient period, $\langle\epsilon(t)\rangle$ converges to
zero as $k/\sqrt{t}$.  Such behavior
is shown in Fig.~\ref{figerror}(a) where $\langle\epsilon(t)\rangle\sqrt{t}$ is
plotted against the simulation time for three values of $\Delta
T$. This is clearly at variance with standard metadynamics in which
the error does not to converge to zero during a single simulation
\cite{laio+05jpcb,buss+06prl}.  The behavior of the present scheme is
consistent with an error analysis done on a simulation performed at a
constant bias.  In Fig.~\ref{figerror}(b)
we study the dependence of $k=\lim_{t\rightarrow\infty} \langle\epsilon(t)\rangle\sqrt{t}$ on $\Delta T$
as a way of optimizing the choice of $\Delta T$.  In this case the
optimal choice is close to $\Delta T= 1200 K$ resulting in a sampling
temperature for the collective variables of $T + \Delta T= 1500 K$,
which is of the order of magnitude of the barrier height. Its actual
value may depend on the $s$ relaxation times and on the area one wishes
to explore.

We discuss now the role of $\omega$, the initial
deposition rate which we relate to the time constant $\tau_{\text{B}}= \frac{\Delta T}{\omega}$
that sets the time scale for the bias evolution.
While in the long time limit $\tau_{\text{B}}$ is irrelevant,
it could affect the transient regime in a non-trivial way.
At constant $\Delta T$, a small $\tau_{\text{B}}$ implies a high initial deposition
rate, thus leading to rapid filling of the wells.
However, if $\tau_{\text{B}}$ is too small relative to the time
necessary to properly average out the transverse degrees of freedom,
the large fluctuations in the initial FES reconstruction need
a longer time to be recovered.

This effect is conveniently investigated by introducing an artificial
model based on the alanine dipeptide FES.  We model the dynamics on
the two-dimensional space $(\Phi,\Psi)$ with a high-friction Langevin
equation driven by the free energy surface $F(\Phi,\Psi)$ and the
diffusion coefficients $D_{\Phi}$, $D_{\Psi}$ determined from the
atomistic simulations.  We shall apply our scheme to calculate the
one-dimensional projection $F(\Phi)$, using a one-dimensional bias on
$\Phi$.  In such a case the relaxation speed of the transverse degree of
freedom $\Psi$ can be tuned by changing 
$D_{\Psi}$, thus mimicking a situation in which the transverse degrees
of freedom are either fast or slow.  As can be seen in Fig.~\ref{figmodel},
in the
fast case, the orthogonal degree of freedom is rapidly averaged out,
resulting in a Markovian dynamics on $\Psi$, and a small $\tau_{\text{B}}$ is
the best choice.  In the slow case, the effective dynamics of $\Phi$
is strongly non-Markovian due to coupling with $\Psi$, and a small
$\tau_{\text{B}}$ is not the best choice since it results in an increase of
the transient time.  However, it is worth noting that the method is
robust and in the range of reported cases, which spans two orders of
magnitude in $\tau_{\text{B}}$ and $D_{\Psi}$, the calculation converged to the
same results on approximately the same time scale.

In conclusion, well-tempered metadynamics
solves the convergence problems of metadynamics and allows
the computational effort to be focused on the physically relevant regions of the
conformational space.  The latter property makes it possible to use
adaptive-bias methods in higher dimensionality cases, thus paving the
way for the study of complex systems where it is difficult to select
{\it a priori} a very small number of relevant degrees of freedom.
The proposed approach can easily be applied to generalizations of
metadynamics based on multiple replicas
\cite{rait+06jpcb,buss+06jacs,pian-laio07jpcb}, and can be extended to
the Wang-Landau algorithm \cite{wang-land01prl}.  

The authors acknowledge Davide Branduardi and Francesco L. Gervasio
for useful discussions.


\begin{thebibliography}{37}
\expandafter\ifx\csname natexlab\endcsname\relax\def\natexlab#1{#1}\fi
\expandafter\ifx\csname bibnamefont\endcsname\relax
  \def\bibnamefont#1{#1}\fi
\expandafter\ifx\csname bibfnamefont\endcsname\relax
  \def\bibfnamefont#1{#1}\fi
\expandafter\ifx\csname citenamefont\endcsname\relax
  \def\citenamefont#1{#1}\fi
\expandafter\ifx\csname url\endcsname\relax
  \def\url#1{\texttt{#1}}\fi
\expandafter\ifx\csname urlprefix\endcsname\relax\def\urlprefix{URL }\fi
\providecommand{\bibinfo}[2]{#2}
\providecommand{\eprint}[2][]{\url{#2}}

\bibitem[{\citenamefont{Gear et~al.}(2002)\citenamefont{Gear, Kevrekidis, and
  Theodoropoulos}}]{gear+02cce}
\bibinfo{author}{\bibfnamefont{C.~W.} \bibnamefont{Gear}},
  \bibinfo{author}{\bibfnamefont{I.~G.} \bibnamefont{Kevrekidis}},
  \bibnamefont{and}
  \bibinfo{author}{\bibfnamefont{C.}~\bibnamefont{Theodoropoulos}},
  \bibinfo{journal}{Comput. Chem. Eng.} \textbf{\bibinfo{volume}{26}},
  \bibinfo{pages}{941} (\bibinfo{year}{2002}).

\bibitem[{\citenamefont{Hummer and Kevrekidis}(2003)}]{humm-kevr03jcp}
\bibinfo{author}{\bibfnamefont{G.}~\bibnamefont{Hummer}} \bibnamefont{and}
  \bibinfo{author}{\bibfnamefont{I.~G.} \bibnamefont{Kevrekidis}},
  \bibinfo{journal}{J.~Chem.~Phys.} \textbf{\bibinfo{volume}{118}},
  \bibinfo{pages}{10762} (\bibinfo{year}{2003}).

\bibitem[{\citenamefont{Torrie and Valleau}(1977)}]{torri-valle77jcp}
\bibinfo{author}{\bibfnamefont{G.~M.} \bibnamefont{Torrie}} \bibnamefont{and}
  \bibinfo{author}{\bibfnamefont{J.~P.} \bibnamefont{Valleau}},
  \bibinfo{journal}{J.~Comput.~Phys.} \textbf{\bibinfo{volume}{23}},
  \bibinfo{pages}{187} (\bibinfo{year}{1977}).

\bibitem[{\citenamefont{Wang and Landau}(2001)}]{wang-land01prl}
\bibinfo{author}{\bibfnamefont{F.}~\bibnamefont{Wang}} \bibnamefont{and}
  \bibinfo{author}{\bibfnamefont{D.~P.} \bibnamefont{Landau}},
  \bibinfo{journal}{Phys.~Rev.~Lett.} \textbf{\bibinfo{volume}{86}},
  \bibinfo{pages}{2050} (\bibinfo{year}{2001}).

\bibitem[{\citenamefont{Darve and Pohorille}(2001)}]{darv-poho01jcp}
\bibinfo{author}{\bibfnamefont{E.}~\bibnamefont{Darve}} \bibnamefont{and}
  \bibinfo{author}{\bibfnamefont{A.}~\bibnamefont{Pohorille}},
  \bibinfo{journal}{J.~Chem.~Phys.} \textbf{\bibinfo{volume}{115}},
  \bibinfo{pages}{9169} (\bibinfo{year}{2001}).

\bibitem[{\citenamefont{Laio and Parrinello}(2002)}]{laio-parr02pnas}
\bibinfo{author}{\bibfnamefont{A.}~\bibnamefont{Laio}} \bibnamefont{and}
  \bibinfo{author}{\bibfnamefont{M.}~\bibnamefont{Parrinello}},
  \bibinfo{journal}{Proc.~Natl.~Acad.~Sci.~U.S.A.}
  \textbf{\bibinfo{volume}{99}}, \bibinfo{pages}{12562} (\bibinfo{year}{2002}).

\bibitem[{\citenamefont{Marsili et~al.}(2006)\citenamefont{Marsili, Barducci,
  Chelli, Procacci, and Schettino}}]{mars+06jpcb}
\bibinfo{author}{\bibfnamefont{S.}~\bibnamefont{Marsili}},
  \bibinfo{author}{\bibfnamefont{A.}~\bibnamefont{Barducci}},
  \bibinfo{author}{\bibfnamefont{R.}~\bibnamefont{Chelli}},
  \bibinfo{author}{\bibfnamefont{P.}~\bibnamefont{Procacci}}, \bibnamefont{and}
  \bibinfo{author}{\bibfnamefont{V.}~\bibnamefont{Schettino}},
  \bibinfo{journal}{J.~Phys.~Chem.~B} \textbf{\bibinfo{volume}{110}},
  \bibinfo{pages}{14011} (\bibinfo{year}{2006}).

\bibitem[{\citenamefont{Micheletti et~al.}(2004)\citenamefont{Micheletti, Laio,
  and Parrinello}}]{mich+04prl}
\bibinfo{author}{\bibfnamefont{C.}~\bibnamefont{Micheletti}},
  \bibinfo{author}{\bibfnamefont{A.}~\bibnamefont{Laio}}, \bibnamefont{and}
  \bibinfo{author}{\bibfnamefont{M.}~\bibnamefont{Parrinello}},
  \bibinfo{journal}{Phys.~Rev.~Lett.} \textbf{\bibinfo{volume}{92}},
  \bibinfo{pages}{170601} (\bibinfo{year}{2004}).

\bibitem[{\citenamefont{Gervasio et~al.}(2005)\citenamefont{Gervasio, Laio, and
  Parrinello}}]{gerv+05jacs}
\bibinfo{author}{\bibfnamefont{F.~L.} \bibnamefont{Gervasio}},
  \bibinfo{author}{\bibfnamefont{A.}~\bibnamefont{Laio}}, \bibnamefont{and}
  \bibinfo{author}{\bibfnamefont{M.}~\bibnamefont{Parrinello}},
  \bibinfo{journal}{J.~Am.~Chem.~Soc.} \textbf{\bibinfo{volume}{127}},
  \bibinfo{pages}{2600} (\bibinfo{year}{2005}).

\bibitem[{\citenamefont{Ensing and Klein}(2005)}]{ensi-klei05pnas}
\bibinfo{author}{\bibfnamefont{B.}~\bibnamefont{Ensing}} \bibnamefont{and}
  \bibinfo{author}{\bibfnamefont{M.~L.} \bibnamefont{Klein}},
  \bibinfo{journal}{Proc.~Natl.~Acad.~Sci.~U.S.A.}
  \textbf{\bibinfo{volume}{102}}, \bibinfo{pages}{6755} (\bibinfo{year}{2005}).

\bibitem[{\citenamefont{Oganov et~al.}(2005)\citenamefont{Oganov,
  Marton{{\'a}}k, Laio, Raiteri, and Parrinello}}]{ogan+05nature}
\bibinfo{author}{\bibfnamefont{A.~R.} \bibnamefont{Oganov}},
  \bibinfo{author}{\bibfnamefont{R.}~\bibnamefont{Marton{{\'a}}k}},
  \bibinfo{author}{\bibfnamefont{A.}~\bibnamefont{Laio}},
  \bibinfo{author}{\bibfnamefont{P.}~\bibnamefont{Raiteri}}, \bibnamefont{and}
  \bibinfo{author}{\bibfnamefont{M.}~\bibnamefont{Parrinello}},
  \bibinfo{journal}{Nature} \textbf{\bibinfo{volume}{438}},
  \bibinfo{pages}{1142} (\bibinfo{year}{2005}).

\bibitem[{\citenamefont{Ishikawa et~al.}(2006)\citenamefont{Ishikawa, Nagara,
  Kusakabe, and Suzuki}}]{ishi+06prl}
\bibinfo{author}{\bibfnamefont{T.}~\bibnamefont{Ishikawa}},
  \bibinfo{author}{\bibfnamefont{H.}~\bibnamefont{Nagara}},
  \bibinfo{author}{\bibfnamefont{K.}~\bibnamefont{Kusakabe}}, \bibnamefont{and}
  \bibinfo{author}{\bibfnamefont{N.}~\bibnamefont{Suzuki}},
  \bibinfo{journal}{Phys.~Rev.~Lett.} \textbf{\bibinfo{volume}{96}},
  \bibinfo{pages}{095502} (\bibinfo{year}{2006}).

\bibitem[{\citenamefont{Nair et~al.}(2006)\citenamefont{Nair, Schreiner, and
  Marx}}]{nair+06jacs}
\bibinfo{author}{\bibfnamefont{N.~N.} \bibnamefont{Nair}},
  \bibinfo{author}{\bibfnamefont{E.}~\bibnamefont{Schreiner}},
  \bibnamefont{and} \bibinfo{author}{\bibfnamefont{D.}~\bibnamefont{Marx}},
  \bibinfo{journal}{J.~Am.~Chem.~Soc.} \textbf{\bibinfo{volume}{128}},
  \bibinfo{pages}{13815} (\bibinfo{year}{2006}).

\bibitem[{\citenamefont{Boero et~al.}(2006)\citenamefont{Boero, Ikeda, Ito, and
  Terakura}}]{boer+06jacs}
\bibinfo{author}{\bibfnamefont{M.}~\bibnamefont{Boero}},
  \bibinfo{author}{\bibfnamefont{T.}~\bibnamefont{Ikeda}},
  \bibinfo{author}{\bibfnamefont{E.}~\bibnamefont{Ito}}, \bibnamefont{and}
  \bibinfo{author}{\bibfnamefont{K.}~\bibnamefont{Terakura}},
  \bibinfo{journal}{J.~Am.~Chem.~Soc.} \textbf{\bibinfo{volume}{128}},
  \bibinfo{pages}{16798} (\bibinfo{year}{2006}).

\bibitem[{\citenamefont{Bussi et~al.}(2006{\natexlab{a}})\citenamefont{Bussi,
  Gervasio, Laio, and Parrinello}}]{buss+06jacs}
\bibinfo{author}{\bibfnamefont{G.}~\bibnamefont{Bussi}},
  \bibinfo{author}{\bibfnamefont{F.~L.} \bibnamefont{Gervasio}},
  \bibinfo{author}{\bibfnamefont{A.}~\bibnamefont{Laio}}, \bibnamefont{and}
  \bibinfo{author}{\bibfnamefont{M.}~\bibnamefont{Parrinello}},
  \bibinfo{journal}{J.~Am.~Chem.~Soc.} \textbf{\bibinfo{volume}{128}},
  \bibinfo{pages}{13435} (\bibinfo{year}{2006}{\natexlab{a}}).

\bibitem[{\citenamefont{Kumar et~al.}(2007)\citenamefont{Kumar, Kalinichev, and
  Kirkpatrick}}]{kuma+07jcp}
\bibinfo{author}{\bibfnamefont{P.~P.} \bibnamefont{Kumar}},
  \bibinfo{author}{\bibfnamefont{A.~G.} \bibnamefont{Kalinichev}},
  \bibnamefont{and} \bibinfo{author}{\bibfnamefont{R.~J.}
  \bibnamefont{Kirkpatrick}}, \bibinfo{journal}{J.~Chem.~Phys.}
  \textbf{\bibinfo{volume}{126}}, \bibinfo{pages}{204315}
  (\bibinfo{year}{2007}).

\bibitem[{\citenamefont{Spiwok et~al.}(2007)\citenamefont{Spiwok,
  Lipovov{{\'a}}, and Kr{{\'a}}lov{{\'a}}}}]{spiw+07jpcb}
\bibinfo{author}{\bibfnamefont{V.}~\bibnamefont{Spiwok}},
  \bibinfo{author}{\bibfnamefont{P.}~\bibnamefont{Lipovov{{\'a}}}},
  \bibnamefont{and}
  \bibinfo{author}{\bibfnamefont{B.}~\bibnamefont{Kr{{\'a}}lov{{\'a}}}},
  \bibinfo{journal}{J.~Phys.~Chem.~B} \textbf{\bibinfo{volume}{111}},
  \bibinfo{pages}{3073} (\bibinfo{year}{2007}).

\bibitem[{\citenamefont{Lee et~al.}(2006)\citenamefont{Lee, Asciutto, Babin,
  Sagui, Darden, and Roland}}]{lee+06jpcb}
\bibinfo{author}{\bibfnamefont{J.-G.} \bibnamefont{Lee}},
  \bibinfo{author}{\bibfnamefont{E.}~\bibnamefont{Asciutto}},
  \bibinfo{author}{\bibfnamefont{V.}~\bibnamefont{Babin}},
  \bibinfo{author}{\bibfnamefont{C.}~\bibnamefont{Sagui}},
  \bibinfo{author}{\bibfnamefont{T.}~\bibnamefont{Darden}}, \bibnamefont{and}
  \bibinfo{author}{\bibfnamefont{C.}~\bibnamefont{Roland}},
  \bibinfo{journal}{J.~Phys.~Chem.~B} \textbf{\bibinfo{volume}{110}},
  \bibinfo{pages}{2325} (\bibinfo{year}{2006}).

\bibitem[{\citenamefont{Huber et~al.}(1994)\citenamefont{Huber, Torda, and van
  Gunsteren}}]{hube+94jcamd}
\bibinfo{author}{\bibfnamefont{T.}~\bibnamefont{Huber}},
  \bibinfo{author}{\bibfnamefont{A.~E.} \bibnamefont{Torda}}, \bibnamefont{and}
  \bibinfo{author}{\bibfnamefont{W.~F.} \bibnamefont{van Gunsteren}},
  \bibinfo{journal}{J. Comput.-Aided Mol. Des.} \textbf{\bibinfo{volume}{8}},
  \bibinfo{pages}{695} (\bibinfo{year}{1994}).

\bibitem[{\citenamefont{Bussi et~al.}(2006{\natexlab{b}})\citenamefont{Bussi,
  Laio, and Parrinello}}]{buss+06prl}
\bibinfo{author}{\bibfnamefont{G.}~\bibnamefont{Bussi}},
  \bibinfo{author}{\bibfnamefont{A.}~\bibnamefont{Laio}}, \bibnamefont{and}
  \bibinfo{author}{\bibfnamefont{M.}~\bibnamefont{Parrinello}},
  \bibinfo{journal}{Phys.~Rev.~Lett.} \textbf{\bibinfo{volume}{96}},
  \bibinfo{pages}{090601} (\bibinfo{year}{2006}{\natexlab{b}}).

\bibitem[{\citenamefont{Laio et~al.}(2005)\citenamefont{Laio, Rodriguez-Fortea,
  Gervasio, Ceccarelli, and Parrinello}}]{laio+05jpcb}
\bibinfo{author}{\bibfnamefont{A.}~\bibnamefont{Laio}},
  \bibinfo{author}{\bibfnamefont{A.}~\bibnamefont{Rodriguez-Fortea}},
  \bibinfo{author}{\bibfnamefont{F.~L.} \bibnamefont{Gervasio}},
  \bibinfo{author}{\bibfnamefont{M.}~\bibnamefont{Ceccarelli}},
  \bibnamefont{and}
  \bibinfo{author}{\bibfnamefont{M.}~\bibnamefont{Parrinello}},
  \bibinfo{journal}{J.~Phys.~Chem.~B} \textbf{\bibinfo{volume}{109}},
  \bibinfo{pages}{6714} (\bibinfo{year}{2005}).

\bibitem[{\citenamefont{Wu et~al.}(2004)\citenamefont{Wu, Schmitt, and
  Car}}]{wu+04jcp}
\bibinfo{author}{\bibfnamefont{Y.}~\bibnamefont{Wu}},
  \bibinfo{author}{\bibfnamefont{J.~D.} \bibnamefont{Schmitt}},
  \bibnamefont{and} \bibinfo{author}{\bibfnamefont{R.}~\bibnamefont{Car}},
  \bibinfo{journal}{J.~Chem.~Phys.} \textbf{\bibinfo{volume}{121}},
  \bibinfo{pages}{1193} (\bibinfo{year}{2004}).

\bibitem[{\citenamefont{Babin et~al.}(2006)\citenamefont{Babin, Roland, Darden,
  and Sagui}}]{babi+06jcp}
\bibinfo{author}{\bibfnamefont{V.}~\bibnamefont{Babin}},
  \bibinfo{author}{\bibfnamefont{C.}~\bibnamefont{Roland}},
  \bibinfo{author}{\bibfnamefont{T.~A.} \bibnamefont{Darden}},
  \bibnamefont{and} \bibinfo{author}{\bibfnamefont{C.}~\bibnamefont{Sagui}},
  \bibinfo{journal}{J.~Chem.~Phys.} \textbf{\bibinfo{volume}{125}},
  \bibinfo{pages}{204909} (\bibinfo{year}{2006}).

\bibitem[{\citenamefont{Min et~al.}(2007)\citenamefont{Min, Liu, Carbone, and
  Yang}}]{min+07jcp}
\bibinfo{author}{\bibfnamefont{D.}~\bibnamefont{Min}},
  \bibinfo{author}{\bibfnamefont{Y.}~\bibnamefont{Liu}},
  \bibinfo{author}{\bibfnamefont{I.}~\bibnamefont{Carbone}}, \bibnamefont{and}
  \bibinfo{author}{\bibfnamefont{W.}~\bibnamefont{Yang}},
  \bibinfo{journal}{J.~Chem.~Phys.} \textbf{\bibinfo{volume}{126}},
  \bibinfo{pages}{194101} (\bibinfo{year}{2007}).

\bibitem[{\citenamefont{Poulain et~al.}(2006)\citenamefont{Poulain, Calvo,
  Antoine, Broyer, and Dugourd}}]{poul+06pre}
\bibinfo{author}{\bibfnamefont{P.}~\bibnamefont{Poulain}},
  \bibinfo{author}{\bibfnamefont{F.}~\bibnamefont{Calvo}},
  \bibinfo{author}{\bibfnamefont{R.}~\bibnamefont{Antoine}},
  \bibinfo{author}{\bibfnamefont{M.}~\bibnamefont{Broyer}}, \bibnamefont{and}
  \bibinfo{author}{\bibfnamefont{P.}~\bibnamefont{Dugourd}},
  \bibinfo{journal}{Phys.~Rev.~E} \textbf{\bibinfo{volume}{73}},
  \bibinfo{pages}{056704} (\bibinfo{year}{2006}).

\bibitem[{\citenamefont{Harju et~al.}(1997)\citenamefont{Harju, Barbiellini,
  Siljam{\"a}ki, Nieminen, and Ortiz}}]{harj+97prl}
\bibinfo{author}{\bibfnamefont{A.}~\bibnamefont{Harju}},
  \bibinfo{author}{\bibfnamefont{B.}~\bibnamefont{Barbiellini}},
  \bibinfo{author}{\bibfnamefont{S.}~\bibnamefont{Siljam{\"a}ki}},
  \bibinfo{author}{\bibfnamefont{R.~M.} \bibnamefont{Nieminen}},
  \bibnamefont{and} \bibinfo{author}{\bibfnamefont{G.}~\bibnamefont{Ortiz}},
  \bibinfo{journal}{Phys.~Rev.~Lett.} \textbf{\bibinfo{volume}{79}},
  \bibinfo{pages}{1173} (\bibinfo{year}{1997}).

\bibitem[{\citenamefont{Spall}(2003)}]{spal03book}
\bibinfo{author}{\bibfnamefont{J.~C.} \bibnamefont{Spall}},
  \emph{\bibinfo{title}{Introduction to stochastic search and optimization}}
  (\bibinfo{publisher}{Wiley}, \bibinfo{address}{Hoboken, New Jersey},
  \bibinfo{year}{2003}).

\bibitem[{\citenamefont{Belardinelli and Pereyra}(2007)}]{bela-pere07pre}
\bibinfo{author}{\bibfnamefont{R.~E.} \bibnamefont{Belardinelli}}
  \bibnamefont{and} \bibinfo{author}{\bibfnamefont{V.~D.}
  \bibnamefont{Pereyra}}, \bibinfo{journal}{Phys.~Rev.~E}
  \textbf{\bibinfo{volume}{75}}, \bibinfo{pages}{046701}
  (\bibinfo{year}{2007}).

\bibitem[{\citenamefont{VandeVondele and Rothlisberger}(2002)\citenamefont{VandeVondele
  and Rothlisberger}}]{vand-roth02jcp}
\bibinfo{author}{\bibfnamefont{J.}~\bibnamefont{VandeVondele}}, \bibnamefont{and}
\bibinfo{author}{\bibfnamefont{U.}~\bibnamefont{Rothlisberger}},
\bibinfo{journal}{J.~Phys.~Chem.~B} \textbf{\bibinfo{volume}{106}},
  \bibinfo{pages}{203} (\bibinfo{year}{2002}).

\bibitem[{\citenamefont{Rosso et~al.}(2002)\citenamefont{Rosso, Min{{\'a}}ry,
  Zhu, and Tuckerman}}]{ross+02jcp}
\bibinfo{author}{\bibfnamefont{L.}~\bibnamefont{Rosso}},
  \bibinfo{author}{\bibfnamefont{P.}~\bibnamefont{Min{{\'a}}ry}},
  \bibinfo{author}{\bibfnamefont{Z.}~\bibnamefont{Zhu}}, \bibnamefont{and}
  \bibinfo{author}{\bibfnamefont{M.~E.} \bibnamefont{Tuckerman}},
  \bibinfo{journal}{J.~Chem.~Phys.} \textbf{\bibinfo{volume}{116}},
  \bibinfo{pages}{4389} (\bibinfo{year}{2002}).

\bibitem[{\citenamefont{Maragliano and Vanden-Eijnden}(2006)}]{mara-vand06cpl}
\bibinfo{author}{\bibfnamefont{L.}~\bibnamefont{Maragliano}} \bibnamefont{and}
  \bibinfo{author}{\bibfnamefont{E.}~\bibnamefont{Vanden-Eijnden}},
  \bibinfo{journal}{Chem.~Phys.~Lett.} \textbf{\bibinfo{volume}{426}},
  \bibinfo{pages}{168} (\bibinfo{year}{2006}).

\bibitem[{\citenamefont{Maragliano et~al.}(2006)\citenamefont{Maragliano,
  Fischer, Vanden-Eijnden, and Ciccotti}}]{mara+06jcp}
\bibinfo{author}{\bibfnamefont{L.}~\bibnamefont{Maragliano}},
  \bibinfo{author}{\bibfnamefont{A.}~\bibnamefont{Fischer}},
  \bibinfo{author}{\bibfnamefont{E.}~\bibnamefont{Vanden-Eijnden}},
  \bibnamefont{and} \bibinfo{author}{\bibfnamefont{G.}~\bibnamefont{Ciccotti}},
  \bibinfo{journal}{J.~Chem.~Phys.} \textbf{\bibinfo{volume}{125}},
  \bibinfo{pages}{024106} (\bibinfo{year}{2006}).

\bibitem[{\citenamefont{Branduardi et~al.}(2007)\citenamefont{Branduardi,
  Gervasio, and Parrinello}}]{bran+07jcp}
\bibinfo{author}{\bibfnamefont{D.}~\bibnamefont{Branduardi}},
  \bibinfo{author}{\bibfnamefont{F.~L.} \bibnamefont{Gervasio}},
  \bibnamefont{and}
  \bibinfo{author}{\bibfnamefont{M.}~\bibnamefont{Parrinello}},
  \bibinfo{journal}{J.~Chem.~Phys.} \textbf{\bibinfo{volume}{126}},
  \bibinfo{pages}{054103} (\bibinfo{year}{2007}).

\bibitem[{\citenamefont{A.~D.~MacKerell
  et~al.}(1998)\citenamefont{A.~D.~MacKerell, Bashford, Bellott, Jr., Evanseck,
  Field, Fischer, Gao, Guo, Ha et~al.}}]{mack+98jpcb}
\bibinfo{author}{\bibfnamefont{J.}~\bibnamefont{A.~D.~MacKerell}},
  \bibinfo{author}{\bibfnamefont{D.}~\bibnamefont{Bashford}},
  \bibinfo{author}{\bibfnamefont{M.}~\bibnamefont{Bellott}},
  \bibinfo{author}{\bibfnamefont{R.~L.~D.} \bibnamefont{Jr.}},
  \bibinfo{author}{\bibfnamefont{J.~D.} \bibnamefont{Evanseck}},
  \bibinfo{author}{\bibfnamefont{M.~J.} \bibnamefont{Field}},
  \bibinfo{author}{\bibfnamefont{S.}~\bibnamefont{Fischer}},
  \bibinfo{author}{\bibfnamefont{J.}~\bibnamefont{Gao}},
  \bibinfo{author}{\bibfnamefont{H.}~\bibnamefont{Guo}},
  \bibinfo{author}{\bibfnamefont{S.}~\bibnamefont{Ha}}, \bibnamefont{et~al.},
  \bibinfo{journal}{J.~Phys.~Chem. B} \textbf{\bibinfo{volume}{102}},
  \bibinfo{pages}{3586} (\bibinfo{year}{1998}).

\bibitem[{\citenamefont{Procacci et~al.}(1997)\citenamefont{Procacci, Darden,
  Paci, and Marchi}}]{ORAC}
\bibinfo{author}{\bibfnamefont{P.}~\bibnamefont{Procacci}},
  \bibinfo{author}{\bibfnamefont{T.~A.} \bibnamefont{Darden}},
  \bibinfo{author}{\bibfnamefont{E.}~\bibnamefont{Paci}}, \bibnamefont{and}
  \bibinfo{author}{\bibfnamefont{M.}~\bibnamefont{Marchi}},
  \bibinfo{journal}{J.~Comp.~Chem.} \textbf{\bibinfo{volume}{18}},
  \bibinfo{pages}{1848} (\bibinfo{year}{1997}).

\bibitem[{\citenamefont{Bussi et~al.}(2007)\citenamefont{Bussi, Donadio, and
  Parrinello}}]{buss+07jcp}
\bibinfo{author}{\bibfnamefont{G.}~\bibnamefont{Bussi}},
  \bibinfo{author}{\bibfnamefont{D.}~\bibnamefont{Donadio}}, \bibnamefont{and}
  \bibinfo{author}{\bibfnamefont{M.}~\bibnamefont{Parrinello}},
  \bibinfo{journal}{J.~Chem.~Phys.} \textbf{\bibinfo{volume}{126}},
  \bibinfo{pages}{014101} (\bibinfo{year}{2007}).

\bibitem[{\citenamefont{Raiteri et~al.}(2005)\citenamefont{Raiteri, Laio,
  Gervasio, Micheletti, and Parrinello}}]{rait+06jpcb}
\bibinfo{author}{\bibfnamefont{P.}~\bibnamefont{Raiteri}},
  \bibinfo{author}{\bibfnamefont{A.}~\bibnamefont{Laio}},
  \bibinfo{author}{\bibfnamefont{F.~L.} \bibnamefont{Gervasio}},
  \bibinfo{author}{\bibfnamefont{C.}~\bibnamefont{Micheletti}},
  \bibnamefont{and}
  \bibinfo{author}{\bibfnamefont{M.}~\bibnamefont{Parrinello}},
  \bibinfo{journal}{J.~Phys.~Chem.~B} \textbf{\bibinfo{volume}{110}},
  \bibinfo{pages}{3533} (\bibinfo{year}{2005}).

\bibitem[{\citenamefont{Piana and Laio}(2007)}]{pian-laio07jpcb}
\bibinfo{author}{\bibfnamefont{S.}~\bibnamefont{Piana}} \bibnamefont{and}
  \bibinfo{author}{\bibfnamefont{A.}~\bibnamefont{Laio}},
  \bibinfo{journal}{J.~Phys.~Chem.~B} \textbf{\bibinfo{volume}{111}},
  \bibinfo{pages}{4553} (\bibinfo{year}{2007}).

\end{thebibliography}
\end{document}